# FIRST STEP IN THEORETICAL APPROACH IN STUDY OF MARS AND TITAN ATMOSPHERES WITH AN INDUCTIVELY COUPLED PLASMA TORCH

André P.[(1)], Clain S. [(2)], Dudeck M.[(3)], Izrar B.[(4)], Rochette D.[(1)], Touzani R.[(2)], Vacher D. [(1)]

*(1) LAEPT, Université Blaise Pascal, Clermont Ferrand, France, pascal.andre@univ-bpclermont.fr*
*(2) LM, Université Blaise Pascal, Clermont Ferrand, France*
*(3) Institut Jean Le Rond d'Alembert, University of Paris 6, France*
*(4) ICARE, CNRS, Université d'Orléans, France.*

## Abstract

To obtain the modelling of an ICP torch that can be used as a test case, we have to determine all the thermodynamic properties and transport coefficients. To calculate the data we have first to determine the composition and the collisions integrals for all the species in the purpose to calculate the transport coefficients. We apply the calculation to the Mars and to the Titan atmosphere compositions. The intensities of the spectral lines are determined versus temperature.

## 1. Introduction

The Mars and Titan atmospheric entries are studied with several ESA test cases. One concerns the ICP torch at atmospheric pressure. The plasma can be considered at thermal and chemical local equilibrium. Some experimental problems (extinction of the plasma) appear with the titan atmosphere ($CH_4$-$N_2$) unlike the Mars atmosphere ($CO_2$-$N_2$). We observe the creation of solid particles certainly graphite and the ICP torch does not work efficiently with the gas similar to the composition of Titan. At the present time no obvious explanations can be provided.

For this purpose, we have begun the modelling of this kind of discharge. It will be useful to understand the energy transfer and the coupling between the induction coils and the plasma gas. The results given by the model will allow us to modify the ICP torch to obtain available plasma with any kind of gas mixtures.

First, we will present a comparison of composition at low temperature (<6000 K). For the temperature included between 3000 K and 15 000 K, we will discuss the composition and monatomic spectral lines for the two considered atmospheres. Secondly, we will describe the basic data to calculate the transport coefficients at thermal equilibrium and at atmospheric pressure. We will discuss the results. Finally, from the thermodynamic properties and transport coefficients, we will describe the hydrodynamic model with the induced current due to magnetic induction.

## 2. Composition, intensities, thermodynamic properties

By using a Gibbs Free Energy minimization method [1, 2], one can obtain the molar fraction versus temperature. The composition of Mars is assumed to be 97% $CO_2$; 3% $N_2$ in molar percentage and for Titan atmosphere 97% $N_2$; 2% $CH_4$; 1% Ar in molar percentage.

For mars atmosphere we take into account 11 monatomic species: C, $C^-$, $C^+$, $C^{++}$, N, $N^+$, $N^{++}$, O, O-, $O^+$, $O^{++}$, 18 diatomic species : $C_2$, $C_2^-$, C2+, CN, CN-, $CN^+$, CO, $CO^-$, $CO^+$, $N_2$, $N_2^-$, $N_2^+$, NO, $NO^-$, $NO^+$, $O_2$, $O_2^-$, $O_2^+$ and 23 polyatomic species: $C_2N$, $C_2N_2$, $C_2O$, $C_3$, $C_3O_2$, $C_4$, $C_4N_2$, $C_5$, CNN, CNO, $CO_2$, $CO_2^-$, $N_2O$, $N_2O_3$, $N_2O_4$, $N_2O_5$, $N_2O$+, $N_3$, NCN, $NO_2$, $NO_2^-$, $NO_3$, $O_3$ and the electrons and the solid phase of carbon: graphite. For Titan atmosphere, we take into account 13 monatomic species: Ar, $Ar^+$, $Ar^{++}$, C, $C^-$, $C^+$, $C^{++}$, H, $H^+$, $H^-$, N, $N^+$, $N^{++}$, 18 diatomic species :$C_2$, $C_2^-$, $C_2^+$, CH, $CH^-$, $CH^+$, CN, $CN^-$, $CN^+$, $H_2$, $H_2^-$, $H_2^+$, $N_2$, $N_2^-$, $N_2^+$, NH, $NH^-$, $H^+$ and 26 polyatomic species: $C_2H$, $C_2H_2$, $C_2H_4$, $C_2N$, $C_2N_2$, $C_3$, $C_4$, $C_4N_2$, $C_5$, $CH_2$, $CH_3$, $CH_4$, CHN, CNN, $H_2N$, $H_2N_2$, $H_3N$, $H_4N_2$, $N_3$, NCN, $H_3$+, $NH_4$+, $C_2H_3$, $C_2H_5$, $C_2$H6, HCCN. The monatomic and diatomic specific thermodynamic properties as enthalpies, chemical potentials are calculated from partitions functions [3, 4]. The formation enthalpies are taken from [8, 9, 10]. The polyatomic species and the carbon solid phase graphite are taken from thermo-chemical tables [8, 9, 10].

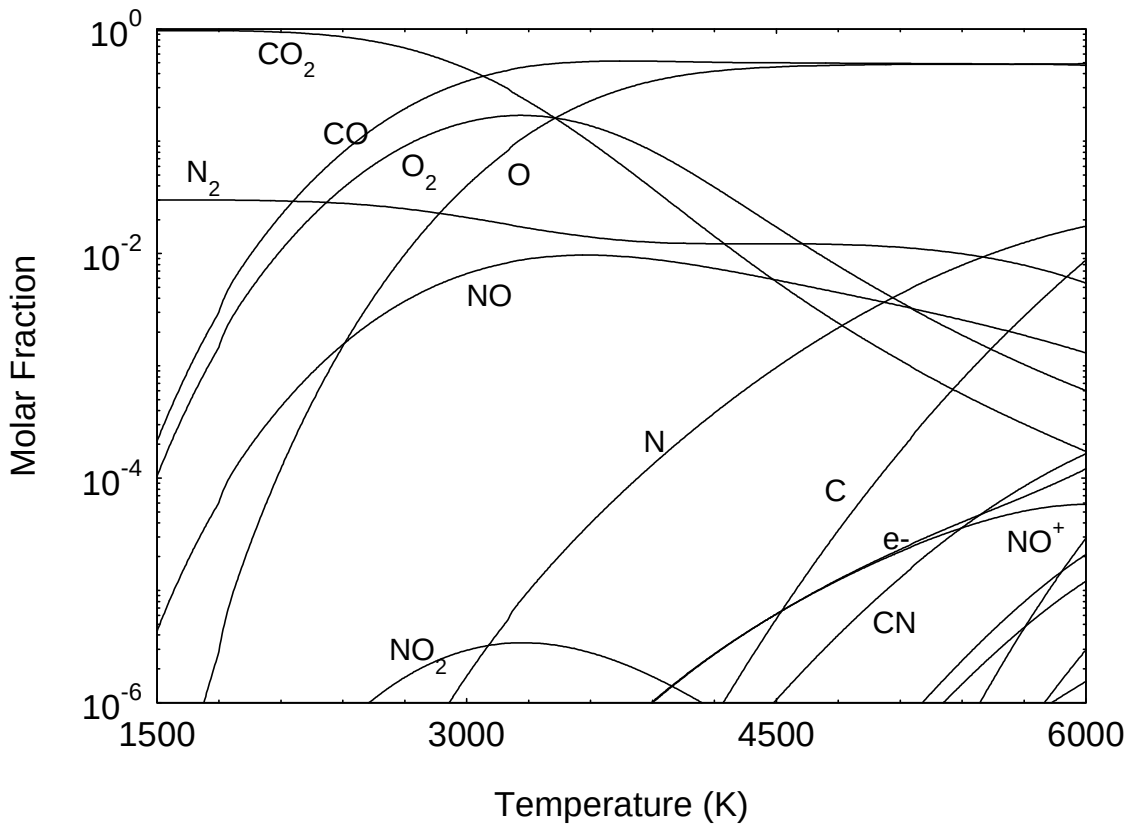

**Figure 1:** Molar fractions of chemical species versus temperature calculated at $10^5$ Pa for Mars atmosphere.

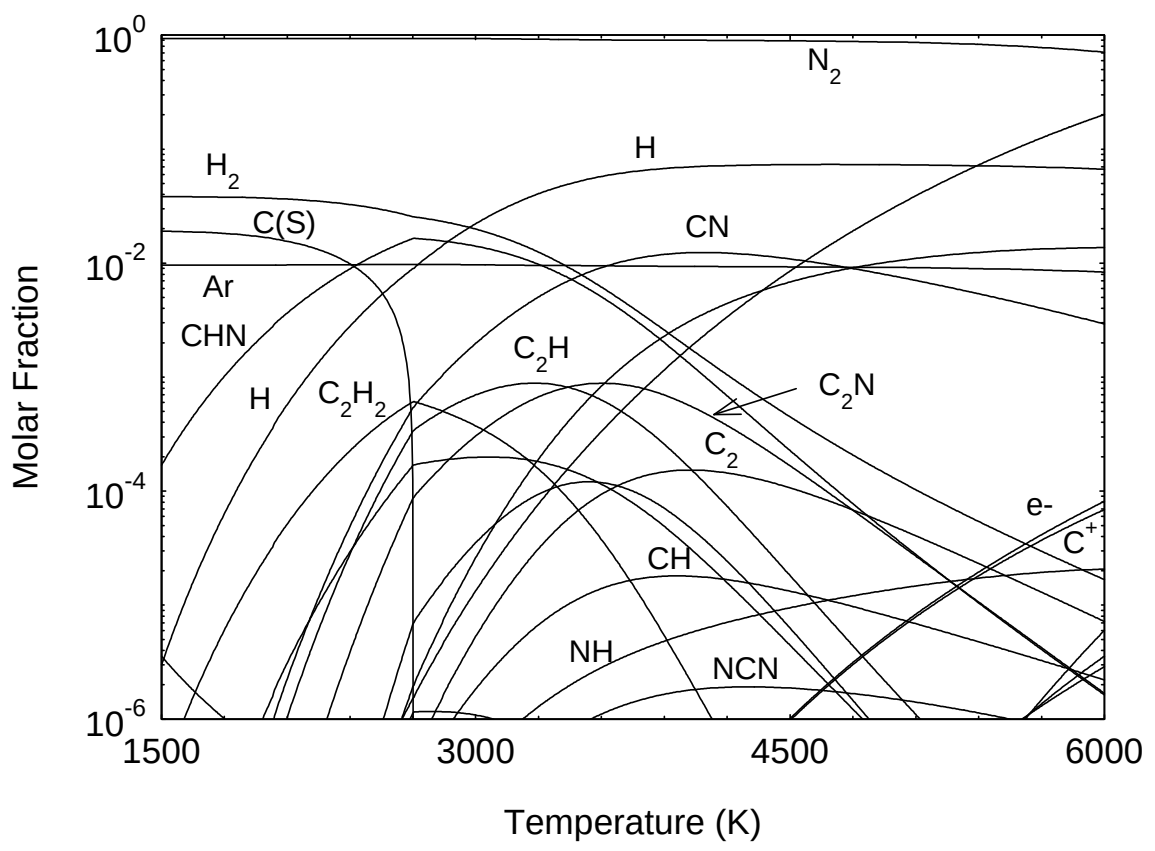

**Figure 2:** Molar fractions of chemical species versus temperature calculated at $10^5$ Pa for Titan atmosphere.

Figure 1 shows the molar fraction versus temperature for Mars atmosphere. We observe that in this temperature range the main chemical species are the $CO_2$ between 1500K and 3000K and CO and O between 3000 K and 6000 K. The electronic neutrality is made between $NO^+$ and $e^-$. The Figure 2 shows the molar fraction versus temperature for Titan atmosphere. We observe that in this temperature range the $CH_4$ molecule is completely dissociated and the main chemical species is $N_2$ that does not dissociate in this temperature range. The electronic neutrality is made between $e^-$ and $C^+$. One main characteristic is the presence of the carbon solid phase: graphite. During the work of ICP torch one can observe the suit formation in the set up. This suit could contain hydrogen.

Since the classical methods to calculate the transport coefficients does not allow to take into account the solid or liquid phase, we choose to study the gas and plasma phase.

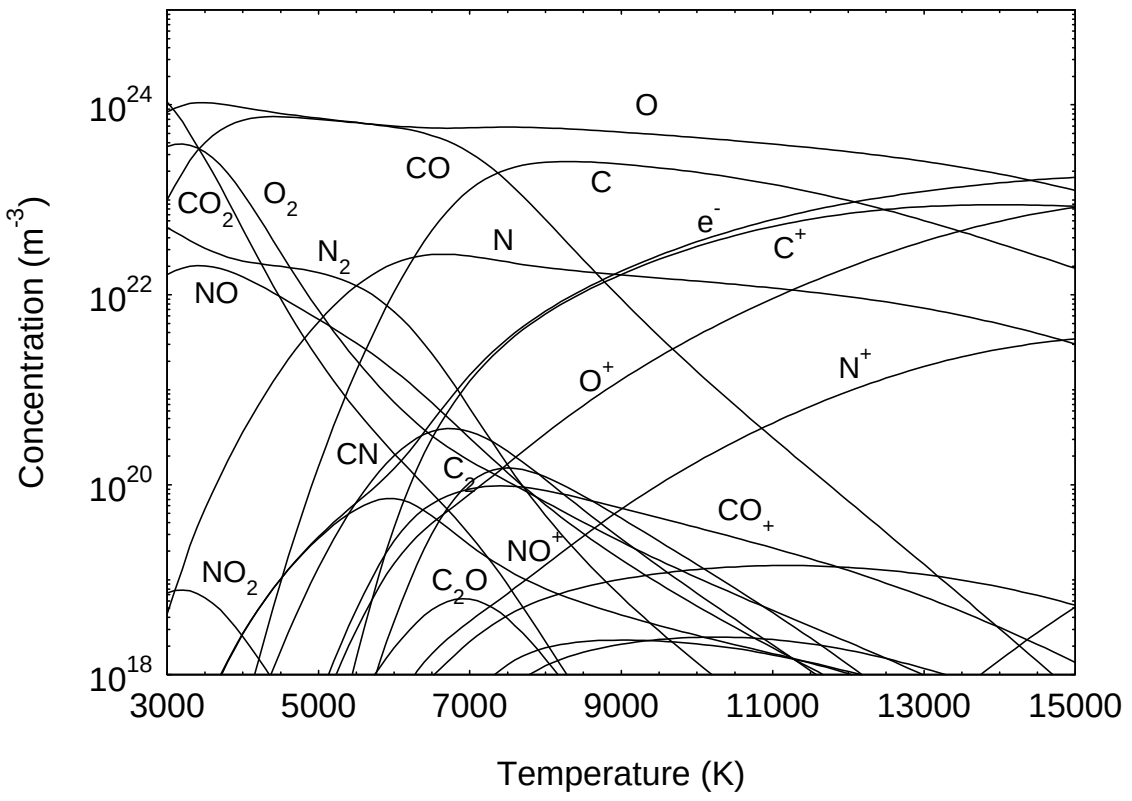

**Figure 3:** Chemical species concentration versus temperature calculated at $10^5$ Pa for Mars atmosphere.

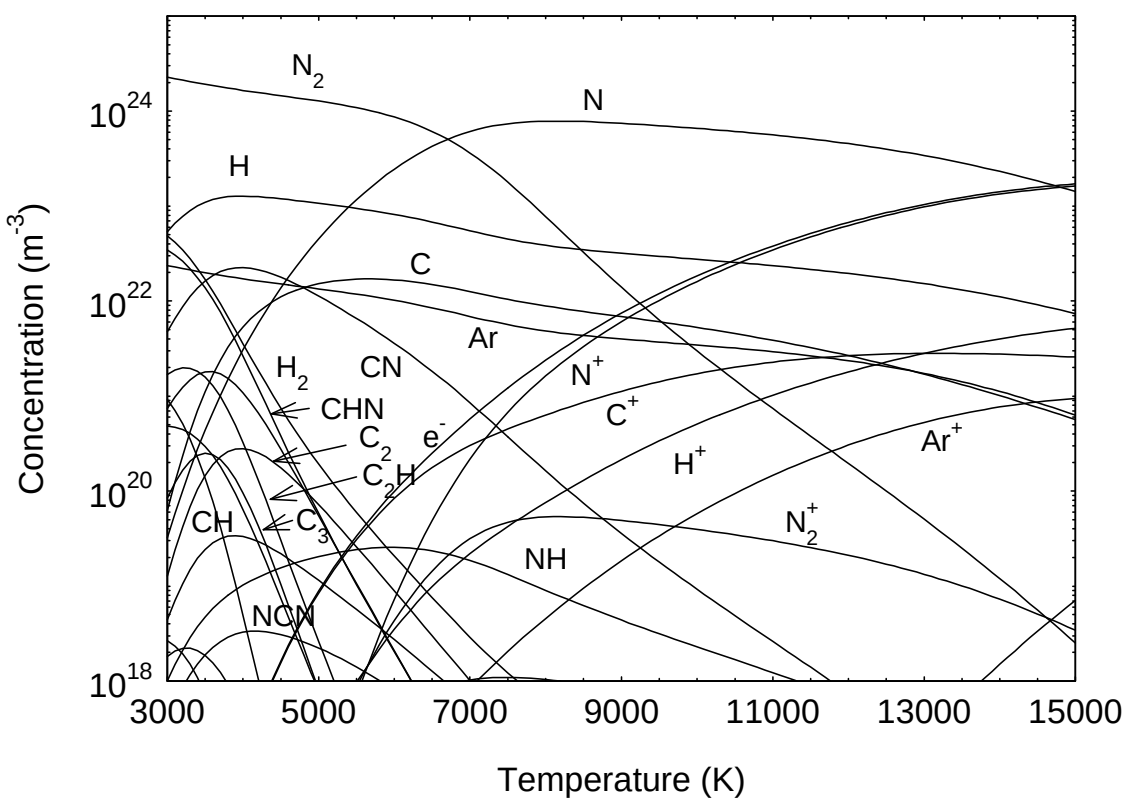

**Figure 4:** Chemical species concentration versus temperature calculated at $10^5$ Pa for Titan atmosphere.

In figure 3, the chemical species versus temperature are shown for the Mars atmosphere. The main chemical species are CO and O for the temperature between 3000K and 7300K and C and O for the higher temperature. The dissociation of CO occurs at a temperature of 7400K. The electrical neutrality is made between $NO^+$ and $e^-$ until 6100 K and between $C^+$ and $e^-$ for higher temperature. In figure 4, the chemical species versus temperature are shown for the Titan atmosphere. The main chemical species are $N_2$ until 6700 K and N for higher temperature. $N_2$ dissociates in N at a temperature around 6700 K. The electrical neutrality is made between $C^+$ and $e^-$ until a temperature of 7100 K and between $N^+$ and $e^-$ for higher temperature. From the composition calculation, assuming a Boltzman distribution one can calculate the intensities of monatomic of spectral lines. We choose to study the triplet oxygen spectral line 777.19, 777.42, 777.54 nm and the carbon spectral line CI 247.86 nm. These spectral lines have been already observed in ICP torch by several authors [11, 12].

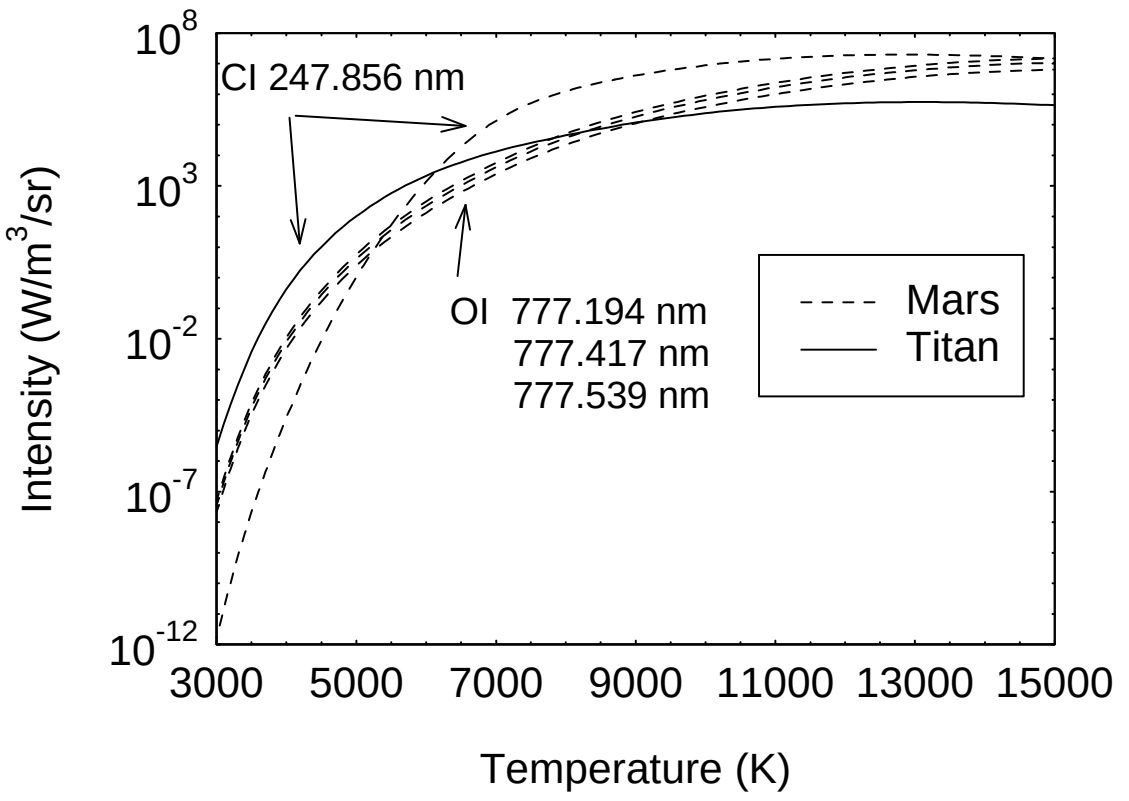

**Figure 5:** Intensities for the considered spectral lines versus temperature for the two atmospheres Mars and Titan.

One interesting feature is the crossing between the Oxygen spectral lines and carbon spectral line around a temperature of 5400 K. This fact allows the experimentalist to roughly evaluate the temperature range below 5400 K or above only by the observance of the spectra.

The density and the internal energy can be determined from the composition and from the specific thermodynamic of each chemical species. These two properties are needed for the ICP torch modeling (§4).

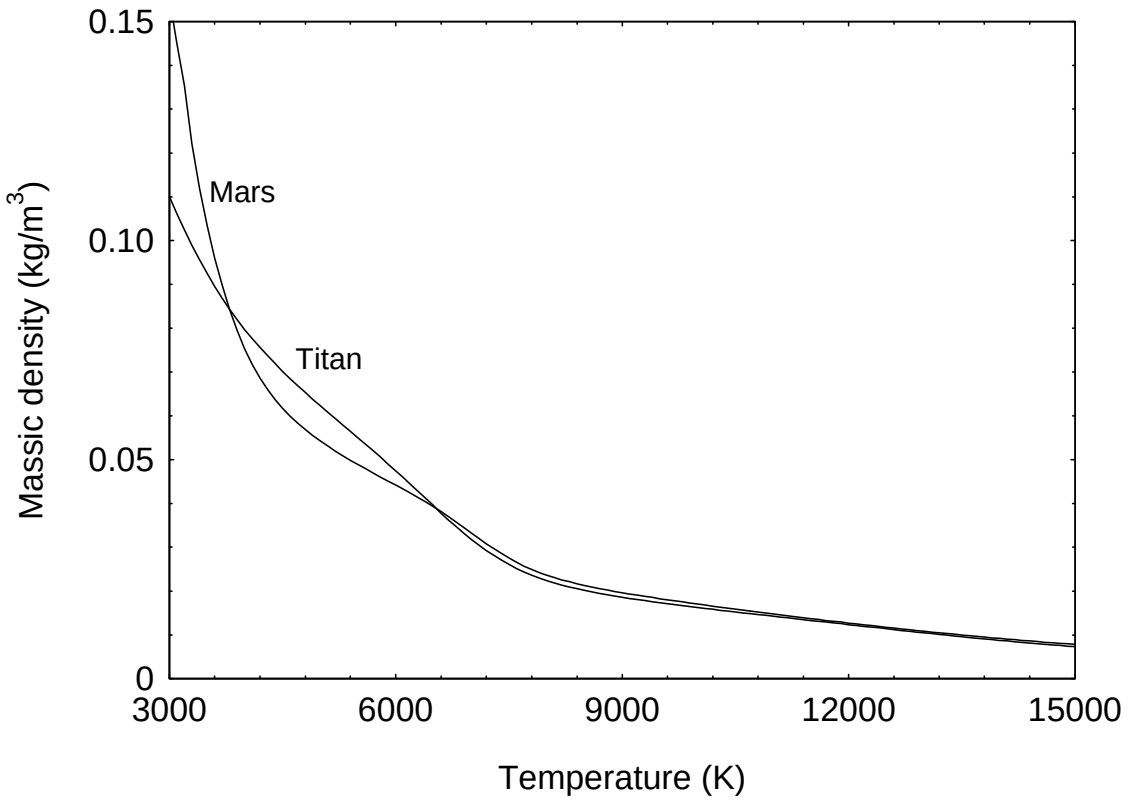

**Figure 6:** Density versus temperature for the two considered atmospheres.

Figure 6 shows the density for the two atmospheres. The density depends on the Dalton law and consequently to the number of chemical species and their mass for a given temperature. That is why when the dissociations have been done the densities are similar i.e. for temperature higher than 7000 K. For the temperature lower than 4000 K the mass of a molecule of $CO_2$ is higher than a molecule of $N_2$ so the density of Mars atmosphere is higher. For intermediate temperature between 4000K and 7000K, the mass of an atom of oxygen is lower than a molecule of $N_2$. So the contribution to the density is lower.

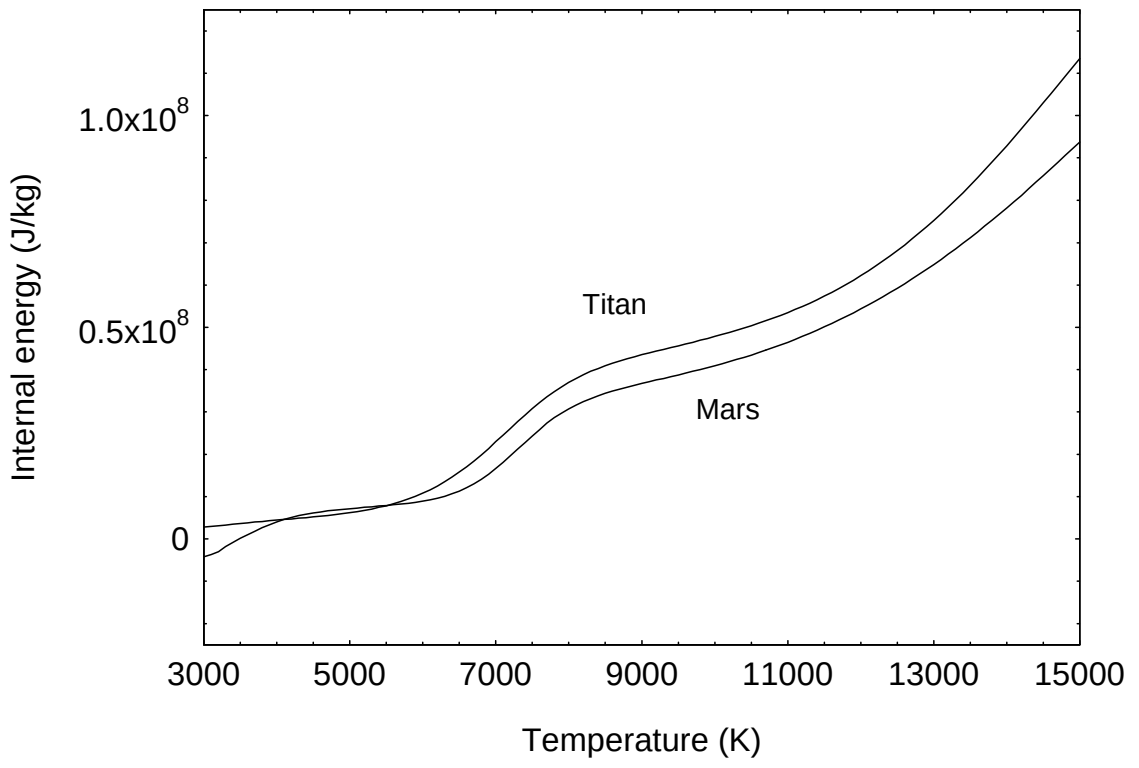

**Figure 7:** Internal energy versus temperature for the two considered atmospheres.

Figure 7 shows the calculated internal energy. The internal energy is calculated from the summation of specific internal energy of each chemical species taking into account the formation enthalpies and the lowering due to the Debye Hückel corrections. We observe some change in the slope of the curves. For Mars successively around 3200K and 7000K that correspond to the dissociation of $CO_2$ and CO. For Titan, around 7000 K that corresponds to the dissociation of $N_2$.

## 3. Transports coefficients

To determine the transport coefficients, we use the well-known solution of the Boltzmann equation due to Chapmann and Enskog [13, 14]. This assumes two-body interactions between chemical species. This interaction can be described by a potential interaction between two particles. By successive integrations of these potentials, we obtain collision integrals which are the basic data of the transport coefficients. The potentials depend on the type of molecules that interact [14, 15, 16]:

- Neutral-neutral molecules
- Charged-neutral molecules
- Charged-charged molecules
- Electron-neutral molecules

For Neutral and Neutral collisions, a Lennard Jones potential are used that allows obtaining of all the potentials for the collisions between neutral molecules with the help of combination rules. For the collisions between neutral and charged molecules a non elastic process namely charged transfer has to be taken into account. For the elastic processes we consider a dipole potential. For the collision between the charged molecules we consider a Coulomb potential shielded by a Debye length. For collisions between electron and neutral species, the collision integrals are obtained from the literature or from the polarisability of neutral species.

Thus the data base to obtain all the collisions integral from the potential are determined.

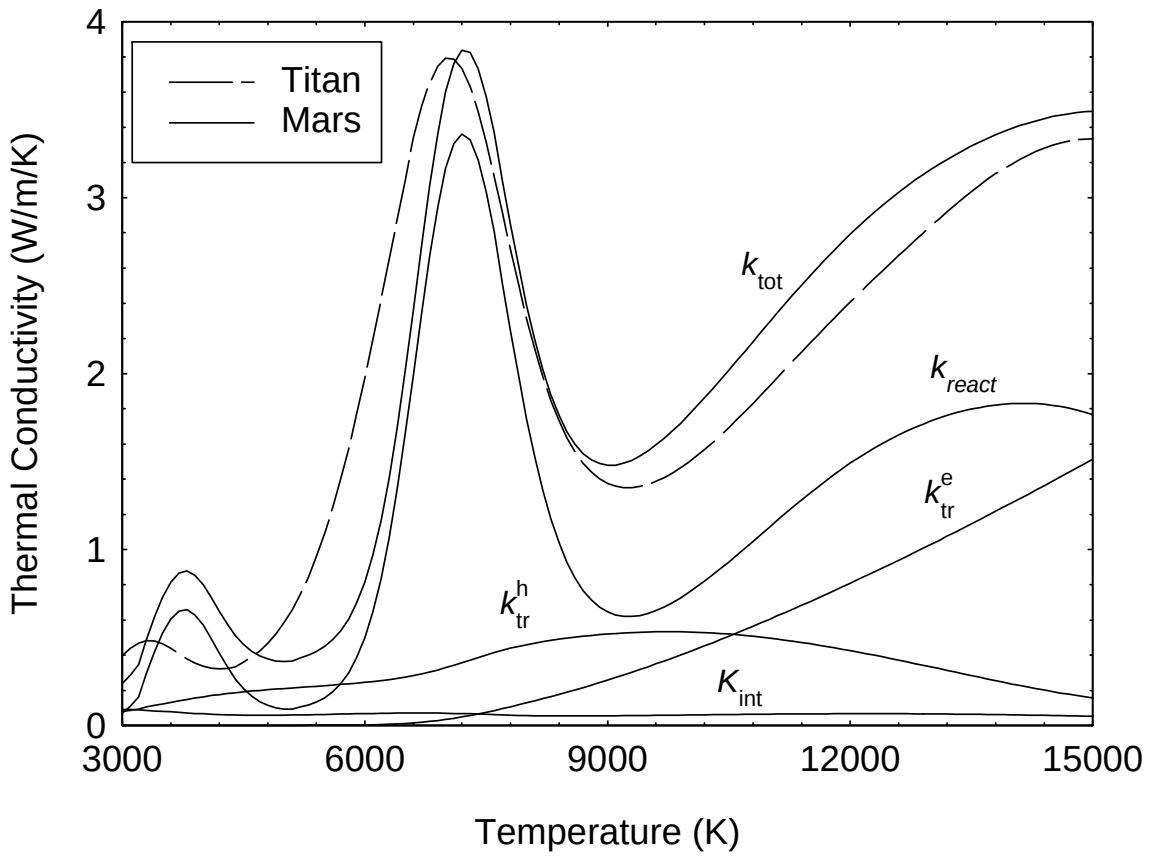

**Figure 8:** Thermal conductivities versus temperature for the two considered atmospheres.

In figure 8, we plot the thermal conductivities for the two considered pressures and for the two considered plasmas. The total thermal conductivity $k_{tot}$ can be separated into four terms with a good accuracy [17, 18]: $k_{tot} = k_{tr}^{e} + k_{tr}^{h} + k_{\text{int}} + k_{react}$ where $k_{tr}^{e}$ is the translational thermal conductivity due to the electrons, $k_{tr}^{h}$ the translational thermal conductivity due to the heavy species particles, $k_{\text{int}}$ the internal thermal conductivity and $k_{react}$ the chemical reaction thermal conductivity. The peak appearing in the thermal conductivity can be associated to the chemical reactions. For Mars atmosphere, we can associate successively the dissociation of $CO_2$, the dissociation of CO and ionisation for temperature around 14 000 K. For titan atmosphere we can associate successively the dissociation of $H_2$, the dissociation of $N_2$ and the ionisation around a temperature of 14000 K.

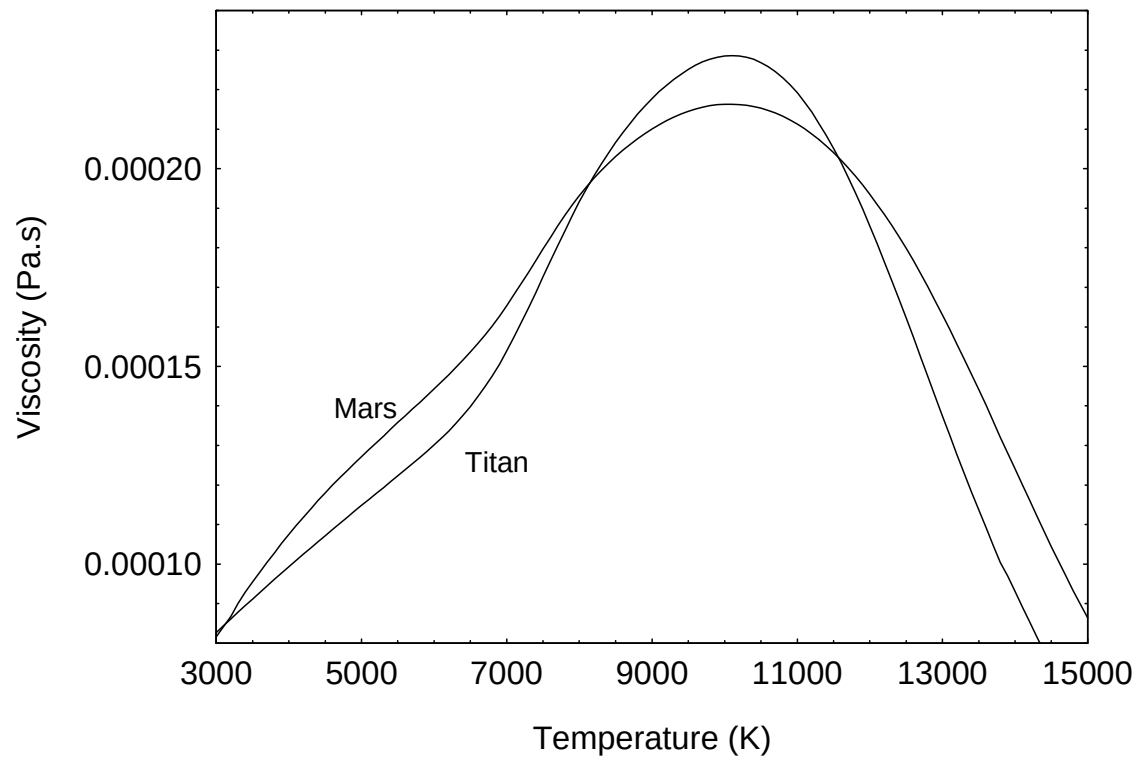

**Figure 9:** Viscosities versus temperature for the two considered atmospheres.

In figure 9, we plot the viscosity versus temperature for the two considered plasmas for Mars and Titan atmospheres. The viscosity for Titan is higher for the temperature between 8000 K and 11500 K than the viscosity for Mars. And its viscosity is lower for the other temperatures in the considered temperature rang 3000 K to 15000 K. This fact has to be confirmed with better potentials as Hulburt Hirschfelder potentials and with better combining rules [19, 20]. Nevertheless, the viscosity of Mars being higher than the viscosity of Titan atmosphere in the temperature range of 3000 to 8000 K, we can assume that the plasma stay a longer time between the coils and then the energy transfer is better for Mars atmosphere for a given temperature. This conclusion has to be confirmed by the complete modelling of the ICP torch.

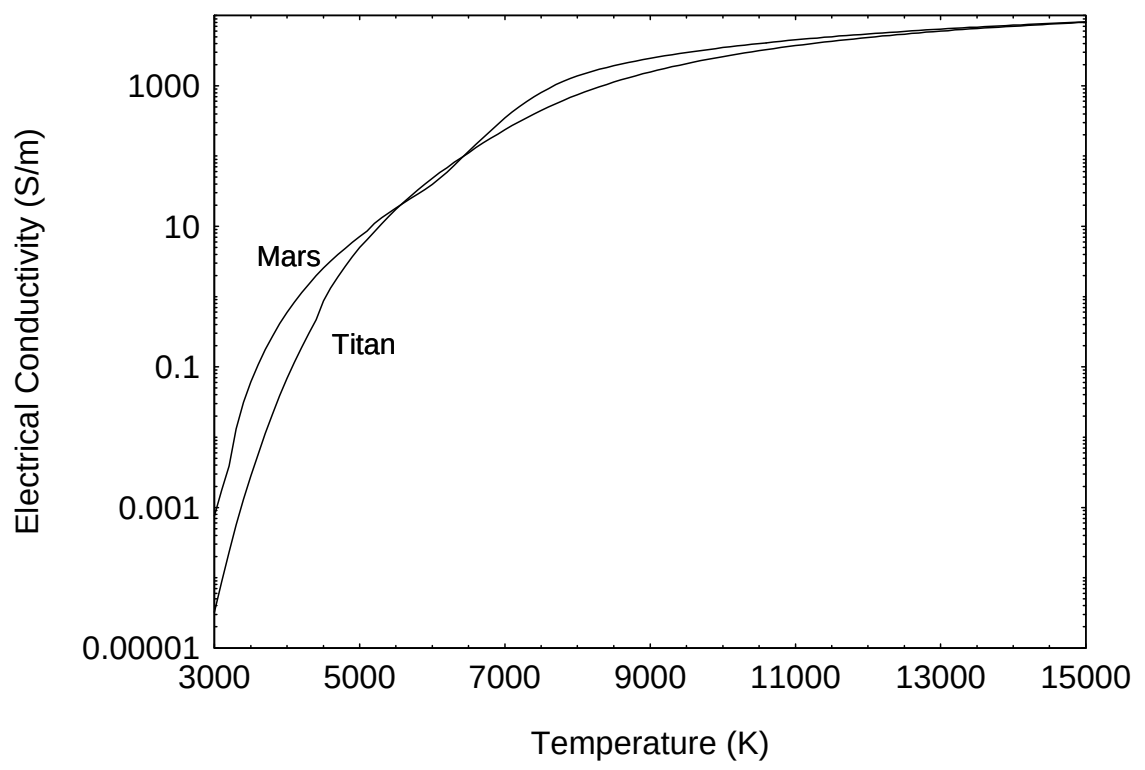

**Figure 10:** Electrical conductivities versus temperature for the two considered atmospheres.

In figure 10, we present the electrical conductivities for the two considered plasma. For Mars atmosphere, at lower temperature below 5000 K, the main collisions are made between electrons and $CO_2$, CO, $O_2$ and O for the temperature between 5000 K and 7500 K the main collisions are made between the electrons and CO and O for the higher temperature between 7500 and 12 000 K between electrons and C and O and between charged particles for higher temperature. For Mars atmosphere, at lower temperature below 7500 K between electrons and $N_2$, between 7500 K and around 13000K between electrons and the monatomic nitrogen N and between electrons and charged particles for higher temperature. We have to notice that we do not take the conductivity of ionic particles in the calculation.

## 4. A mathematical model for inductive plasma torches

We consider a plasma torch device constituted of electrically conducting domains denoted by $\Omega$ and an insulating part $\Omega'$ such that $IR^3 = \Omega \cup \Omega'$. The conducting domain represents the wires which supply the torch the induction coil and the plasma itself considered as a conducting fluid where $\Omega_F$ represents the plasma domain. The insulating part is constituted of the gas and any non-metallic solid part of the set-up.

Mathematical modelling of inductive plasma torches involves a coupling of fluid flow with electromagnetic processes. The Navier-Stokes equations for compressible fluid are considered to model the fluid dynamic while the Maxwell equations are used to describe the electromagnetic fields.

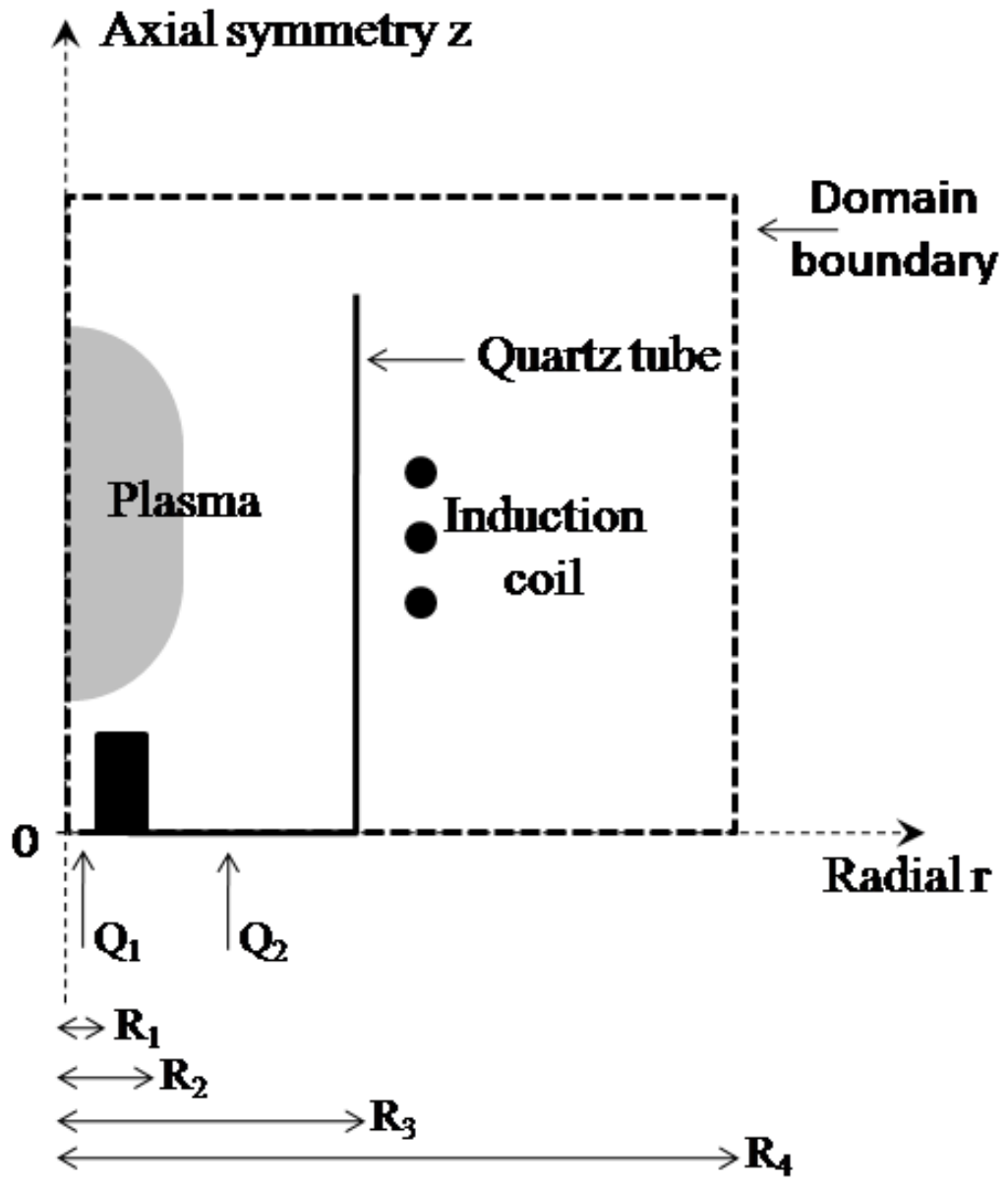

**Figure 11:** Schematic illustration of the plasma torch.

In order to reduce the complexity of the model, we introduce the following assumptions:
- Axisymmetry geometry of the plasma torch.
- Local thermodynamic equilibrium conditions for the plasma.
- Optically thin plasma.
- Laminar fluid flow.
- Negligible viscous dissipation.
- Electromagnetic wave propagation is assumed to be negligible which is motivated by the wave length. Then we resort to the so-called eddy current equations by dropping the displacement current term.
- We assume a given time harmonic source current, i.e., all electromagnetic vector fields have the form $\boldsymbol{J}(x,t)=\mathrm{Re}\left(\boldsymbol{J}(x)e^{i\omega t}\right)\quad x\in IR^3,\ t>0$, where $\omega$ is the angular frequency. The quasi-stationary field $\boldsymbol{J}(x)$ is then complex-valued.

For the electromagnetic part let us denote by ***B, H, J***, ***E*** the magnetic induction, magnetic field, current density and electric field. Assuming that the fields are time harmonic, we have the set of equations:

$$\nabla\times\boldsymbol{H}=\boldsymbol{J}, \qquad (1)$$

$$i\omega\boldsymbol{B}+\nabla\times\boldsymbol{E}=0, \qquad (2)$$

$$\boldsymbol{B}=\mu_0\,\boldsymbol{H}, \qquad (3)$$

$$\boldsymbol{J}=\sigma(\boldsymbol{E}+\boldsymbol{v}\times\boldsymbol{B})+\boldsymbol{J}_0, \qquad (4)$$

where ***v*** is the fluid velocity extended by 0 outside the fluid region, $\boldsymbol{J}_0$ is the given source current density representing the exterior electrical supply, $\mu_0$ is the magnetic permeability of the free space and $\sigma$ is the electric conductivity. Equations (1)-(4) are valid in the whole space in which we distinguish the electric conducting domains denoted by $\Omega=\Omega_1\cup\ldots\Omega_N$ from the free space denote by Ω’.

In the fluid, the electric conductivity depends on the enthalpy $h$ of the fluid where we write $\sigma=\sigma(h)$. The main difficulty is the null conductivity in the domains outside of the plasma so we have to consider a formulation of equations (1)-(4) such that the situation $\sigma=0$ does not lead to a degenerated model producing disastrous numerical approximations. To avoid this drawback, we introduce a mathematical formulation based on the electric field ***E*** defined on the whole space:

$$\nabla\times(\mu_0^{-1}\nabla\times\boldsymbol{E})-\sigma\boldsymbol{v}\times\nabla\times\boldsymbol{E}+i\omega\sigma\boldsymbol{E}=-i\omega\boldsymbol{J}_0 \quad \text{in } IR^3 \qquad (5)$$

$$\boldsymbol{E}(x)=O\left(|x|^{-1}\right) \quad \text{for } |x|\to\infty \qquad (6)$$

**Remark 1.** It is possible to reformulate the equations using the two subdomains Ω and Ω’. To this end, let Γ stand for the boundary of Ω and let [.] stand for the jump of a function through Γ. Using interface conditions that derive from equations (1)-(4), equation (5) can be written as well:

$$\nabla\times(\mu_0^{-1}\nabla\times\boldsymbol{E})-\sigma\boldsymbol{v}\times\nabla\times\boldsymbol{E}+i\omega\sigma\boldsymbol{E}=-i\omega\boldsymbol{J}_0 \quad \text{in } \Omega \qquad (7)$$

$$\nabla\times(\mu_0^{-1}\nabla\times\boldsymbol{E})=0 \quad \text{in } \Omega' \qquad (8)$$

$$[\boldsymbol{E}\times\boldsymbol{n}]=\left[\mu_0^{-1}\nabla\times\boldsymbol{E}\times\boldsymbol{n}\right]=0 \quad \text{on } \Gamma \qquad (9)$$

$$\boldsymbol{E}(x)=O\left(|x|^{-1}\right) \quad \text{for } |x|\to\infty \qquad (10)$$

where the source current density is supported by the conducting domain Ω.

**Remark 2.** Since the fluid motion contribution $\sigma\boldsymbol{v}\times\nabla\times\boldsymbol{E}$ is negligible with respect to $i\omega\sigma\boldsymbol{E}$ we assume that the fluid motion does not modify the electric current and we neglect the convective term involving the velocity field ***v***.

We now assume that the conductors have the symmetry of rotation property with respect to the Oz axis, so we introduce the cylindrical coordinates $(\mathrm{r},\theta,\mathrm{z})$ using $\theta$-invariance and assume that the source current $\boldsymbol{J_0}$ has null radial and axial components, that is, $\boldsymbol{J}_0=\boldsymbol{J}_0(\mathrm{r},\mathrm{z})e^{i\omega t}=J_\theta(\mathrm{r},\mathrm{z})e^{i\omega t}\,\boldsymbol{e}_\theta$.

Then one can show that the electric field enjoys also the same property and that its azimuthal component $\boldsymbol{E}_\theta$ satisfies the equation:

$$-\frac{\partial}{\partial r}\left(\frac{1}{r}\frac{\partial}{\partial r}(rE_\theta)\right)-\frac{\partial^2E_\theta}{\partial z^2}+i\omega\mu_0\sigma E_\theta=-i\omega\mu_0J_\theta. \qquad (11)$$

The fluid flow equations involve the computation of two source terms: the Lorentz force in the momentum equations and the Joule density in the energy equation.

- The Lorentz force density is given by

$$f_{Lorentz}=\frac{1}{2}(\boldsymbol{J}\times\overline{\boldsymbol{B}})=-\frac{i}{2\omega}\mathrm{Re}(\sigma E_\theta+J_\theta)\left(\frac{\partial\overline{E}_\theta}{\partial z}\boldsymbol{e}_z+\frac{1}{r}\frac{\partial}{\partial r}(r\overline{E}_\theta)\boldsymbol{e}_r\right).$$

- The Joule density source term is given by

$$S_{Joule}=\frac{1}{2}\mathrm{Re}(\boldsymbol{J}.\overline{\boldsymbol{E}})=\frac{1}{2}\mathrm{Re}(\sigma E_\theta+J_\theta)\overline{E}_\theta.$$

For the hydrodynamic part, the plasma flow is modelled by the compressible Navier-Stokes equations supplemented by the state equation for the fluid constituting the plasma. The fluid flow is driven by the Lorentz force while induction heating generates a Joule effect term in the energy conservation equation. We obtain the following set of equations written under the conservative form in cylindrical coordinates:

$$\frac{\partial(rU)}{\partial t}+\frac{\partial(rF_r(U))}{\partial r}+\frac{\partial(rF_z(U))}{\partial z}=\frac{\partial(rG_r(U))}{\partial r}+\frac{\partial(rG_z(U))}{\partial z}+S(U) \quad (12)$$

where $U$ is the conservative variables vector defined by

$$U=\begin{pmatrix}\rho\\ \rho u_r\\ \rho u_z\\ \rho u_\theta\\ e\end{pmatrix}.$$

$F_r(U)$ and $F_z(U)$ represent the convective fluxes given by

$$F_r(U)=\begin{pmatrix}\rho u_r\\ \rho u_r^2+P\\ \rho u_z u_r\\ \rho u_\theta u_r\\ u_r(e+P)\end{pmatrix},\ F_z(U)=\begin{pmatrix}\rho u_z\\ \rho u_r u_z\\ \rho u_z^2+P\\ \rho u_\theta u_z\\ u_z(e+P)\end{pmatrix}.$$

$G_r(U)$ and $G_z(U)$ represent the viscous fluxes given by

$$G_r(U)=\begin{pmatrix}0\\ \tau_{rr}\\ \tau_{rz}\\ \tau_{r\theta}\\ q_r\end{pmatrix},\ G_z(U)=\begin{pmatrix}0\\ \tau_{rz}\\ \tau_{zz}\\ \tau_{\theta z}\\ q_z\end{pmatrix}.$$

The source term $S(U)$ is given by

$$S(U)=\begin{pmatrix}0\\ \rho u_\theta^2+P-\tau_{\theta\theta}+rf_r\\ rf_z\\ -\rho u_\theta u_r+\tau_{r\theta}\\ rS_{Joule}-rP_{Rad}\end{pmatrix}.$$

Here, $\rho$ is the fluid mass density, $P$ is the pressure, $e$ is the total energy per unit volume, $u_r$, $u_z$, $u_\theta$ are respectively the radial, axial and tangential velocity components, $P_{Rad}$ is the plasma radiation losses and $S_{Joule}$ is the Joule heating rate.

The components of the viscous tensor are defined by:

$$\tau_{rr}=-\frac{2}{3}\mu\left(\frac{\partial u_r}{\partial r}+\frac{\partial u_z}{\partial z}+\frac{u_r}{r}\right)+2\mu\frac{\partial u_r}{\partial r},$$
$$\tau_{rz}=\tau_{zr}=\mu\left(\frac{\partial u_r}{\partial z}+\frac{\partial u_z}{\partial r}\right),$$
$$\tau_{zz}=-\frac{2}{3}\mu\left(\frac{\partial u_r}{\partial r}+\frac{\partial u_z}{\partial z}+\frac{u_r}{r}\right)+2\mu\frac{\partial u_z}{\partial z},$$
$$\tau_{r\theta}=\mu\left(\frac{\partial u_\theta}{\partial r}-\frac{u_\theta}{r}\right),$$
$$\tau_{\theta z}=\mu\frac{\partial u_\theta}{\partial z},$$
$$\tau_{\theta\theta}=-\frac{2}{3}\mu\left(\frac{\partial u_r}{\partial r}+\frac{\partial u_z}{\partial z}+\frac{u_r}{r}\right)+2\mu\frac{u_r}{r},$$

where $\mu$ is the dynamic viscosity.

The conductive heat flux $\boldsymbol{q}$ is given by:

$$q_r=k\frac{\partial T}{\partial r},\ q_z=k\frac{\partial T}{\partial z},$$

where $k$ is the thermal conductivity.

$f_r$ and $f_z$ are the radial and axial components of the Lorentz force given by:

$$f_r=-\frac{i}{2\omega}\mathrm{Re}(\sigma E_\theta+J_\theta)\left(\frac{1}{r}\frac{\partial}{\partial r}(r\overline{E}_\theta)\right),$$
$$f_z=-\frac{i}{2\omega}\mathrm{Re}(\sigma E_\theta+J_\theta)\frac{\partial\overline{E}_\theta}{\partial z}.$$

In addition, to close the system we add a state equation

$$P=\hat{P}(\rho,\varepsilon)$$

where ε stands for the specific internal energy related to the total energy by

$$e=\rho\varepsilon+\frac{1}{2}\rho|\boldsymbol{u}|^2.$$

.

The boundary conditions for equation (12) are as follows:

- Inlet conditions $z=0$:

  $0<r<R_1:\ u_z=\frac{Q_1}{\pi R_1^2},\ u_r=u_\theta=0,$

  $P=1bar,\ T=298K.$

  $R_2<r<R_3:\ u_z=\frac{Q_2}{\pi(R_3^2-R_2^2)},\ u_r=0,\ u_\theta=cte,$

  $P=1bar,\ T=298K.$

- Axial conditions $r=0$:

  $\frac{\partial u_r}{\partial r}=u_\theta=u_z=\frac{\partial P}{\partial r}=\frac{\partial T}{\partial r}=0.$

- Domain boundary:

$$\frac{\partial u_r}{\partial r}=\frac{\partial u_\theta}{\partial r}=\frac{\partial u_z}{\partial r}=\frac{\partial P}{\partial r}=\frac{\partial T}{\partial r}=0.$$

To avoid difficulties related to unbounded domains, we replace the condition at the infinity for the electric field equation (11) by the condition $E_\theta = 0$ on the boundary of a large box that contains the conductors. In this paragraph, we have presented the equations that has to be resolved soon taking the two set of data one for Mars atmosphere and the second for Titan atmosphere.

## 4. Conclusion

From the calculation of the composition in the polyphasic area we have shown that for Mars atmosphere we have no suit formation unlike the Titan atmosphere. This fact has been readily experimentally observed. From the composition we have calculate the intensity of interesting spectral lines of carbon CI 247.86 nm and the oxygen triplet spectral line 777.19, 777.42, 777.54 nm that can allows the experimentalist to roughly estimate the temperature in the plasma assuming the Boltzman distribution. Furthermore, we have presented all the data needed to model the ICP Torch: internal energy, density, electrical and thermal conductivity and viscosity for Mars atmosphere and Titan atmosphere. The equations of the modelling of the ICP torch have been given. It will be useful to compare the modelling with other modelling [21] test the sensibility and the needed accuracy of transport coefficients and at last compare the radiation power loss with the experimental results and the total modelling spectra.

## Acknowledgements

The authors acknowledge many interesting and useful discussions in the ANR group Rayhen. The authors also acknowledge the starting ESA TRP "Validation of Aerothermochemistry Models for Re-Entry Applications" for the interest in their researches.